\documentclass[conference]{IEEEtran}
\IEEEoverridecommandlockouts
\usepackage{cite}
\usepackage{amsmath,amssymb,amsfonts}
\usepackage{graphicx}
\usepackage{textcomp}
\usepackage{amsmath}
\usepackage{booktabs} 
\usepackage{multirow}
\usepackage{hyperref}       
\usepackage{url}            
\usepackage{booktabs}       
\usepackage{amsfonts}       
\usepackage{nicefrac}       
\usepackage{microtype}      
\usepackage{xcolor}         
\usepackage{algorithm}
\usepackage{algpseudocode}
\usepackage{amsmath}
\usepackage{tikz}
\usepackage{adjustbox}

\usepackage{orcidlink}
\usepackage{comment}
\usepackage{amssymb}
\usepackage{braket}
\usepackage{xcolor}
\def\BibTeX{{\rm B\kern-.05em{\sc i\kern-.025em b}\kern-.08em
    T\kern-.1667em\lower.7ex\hbox{E}\kern-.125emX}}
\begin{document}

\title{HQNN-FSP: A Hybrid Classical-Quantum Neural Network for Regression-Based Financial Stock Market Prediction}

\author{\IEEEauthorblockN{\textbf{Prashant Kumar Choudhary}\textsuperscript{1}, \textbf{Nouhaila Innan}\textsuperscript{2,3}, \textbf{Muhammad Shafique}\textsuperscript{2,3}, \textbf{Rajeev Singh}\textsuperscript{1}
\IEEEauthorblockA{
\textsuperscript{1}Department of Physics, Indian Institute of Technology (BHU), Varanasi, 221005, India\\
\textsuperscript{2}eBRAIN Lab, Division of Engineering, New York University Abu Dhabi (NYUAD), Abu Dhabi, UAE\\
\textsuperscript{3}Center for Quantum and Topological Systems (CQTS), NYUAD Research Institute, NYUAD, Abu Dhabi, UAE\\
Emails: prashantkchoudhary.rs.phy22@iitbhu.ac.in, nouhaila.innan@nyu.edu, \\muhammad.shafique@nyu.edu, rajeevs.phy@iitbhu.ac.in\\
}}}

\maketitle

\begin{abstract}
 Financial time-series forecasting remains a challenging task due to complex temporal dependencies and market fluctuations. This study explores the potential of hybrid quantum-classical approaches to assist in financial trend prediction by leveraging quantum resources for improved feature representation and learning. A custom Quantum Neural Network (QNN) regressor is introduced, designed with a novel ansatz tailored for financial applications. Two hybrid optimization strategies are proposed: (1) a sequential approach where classical recurrent models (RNN/LSTM) extract temporal dependencies before quantum processing, and (2) a joint learning framework that optimizes classical and quantum parameters simultaneously. Systematic evaluation using TimeSeriesSplit, k-fold cross-validation, and predictive error analysis highlights the ability of these hybrid models to integrate quantum computing into financial forecasting workflows. The findings demonstrate how quantum-assisted learning can contribute to financial modeling, offering insights into the practical role of quantum resources in time-series analysis.
\end{abstract}

\section{Introduction}
\begin{figure}[htpb] 
\centering 
\includegraphics[width=\linewidth]{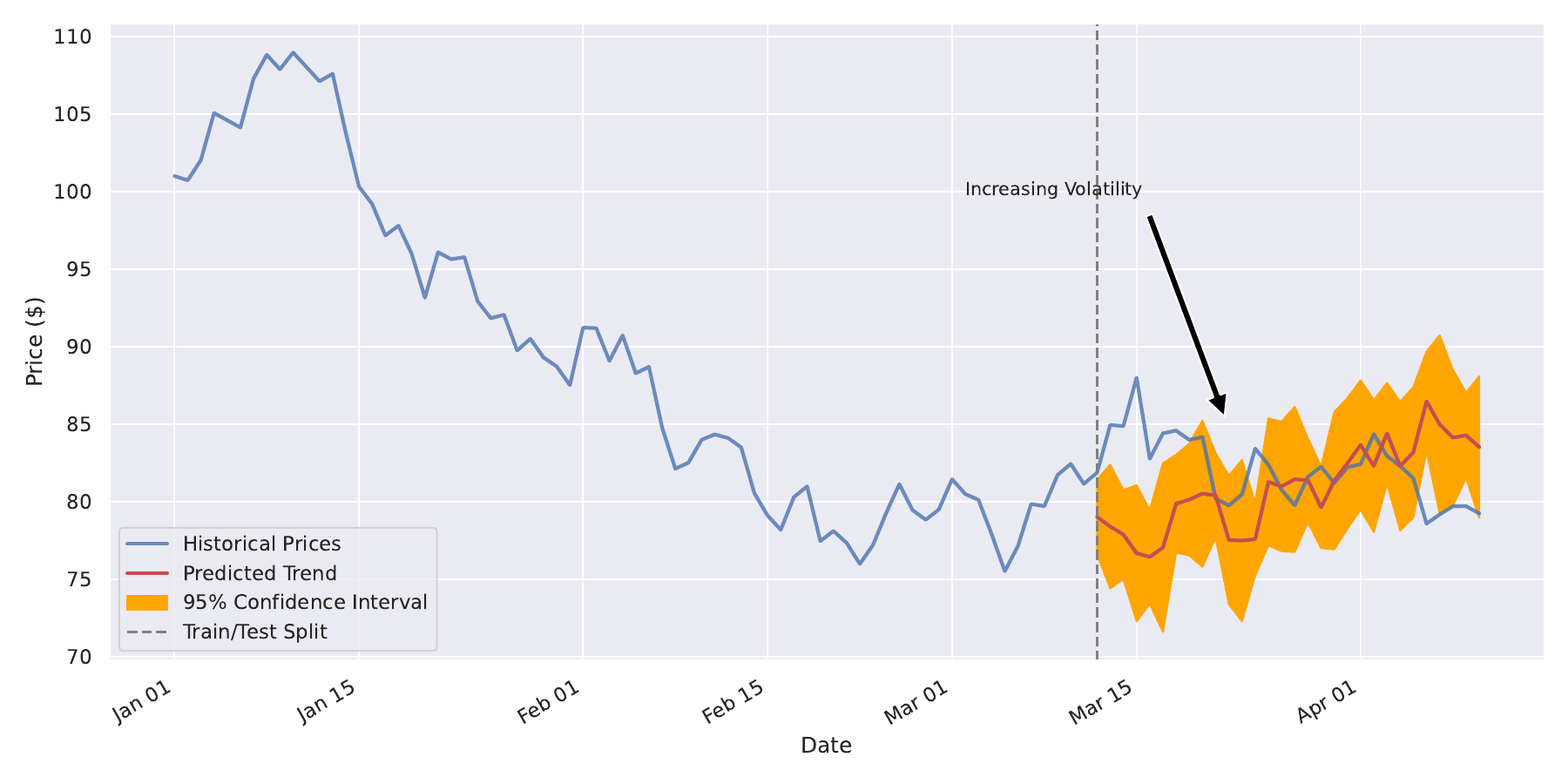} 
\vspace{-0.8cm}
\caption{Stock price forecasting plot displaying historical prices (blue), a predicted trend (red) with a 95\% confidence interval (shaded), and a clear train/test split. The observed increasing volatility highlights the challenges of financial forecasting and underscores the need for robust hybrid quantum-classical models.} 
\label{fig:Financial}
 \end{figure}

Stock market forecasting is a critical area of financial research due to its significant economic implications (see Fig. \ref{fig:Financial}). Accurate predictions enable informed investment decisions, mitigating risks while optimizing returns~\cite{bib17}. However, stock price movements are inherently complex, exhibiting high volatility, non-linearity, and dependencies on numerous macroeconomic factors, geopolitical events, and investor sentiment dynamics~\cite{bib1,bib18,bib24}. Traditional Machine Learning (ML) models, such as Recurrent Neural Networks (RNNs) and Long Short-Term Memory (LSTM) networks, have been extensively used to model temporal dependencies in financial time series~\cite{bib2}. While these methods have shown success, they face several limitations, including overfitting, inadequate generalization in dynamic market conditions, and difficulty capturing long-range dependencies~\cite{bib3,bib4}. These challenges necessitate the exploration of alternative computational paradigms that can better address the intricacies of financial time series~\cite{bib8}.

Quantum Machine Learning (QML) has emerged as a promising approach that integrates Quantum Computing (QC) principles with ML methodologies. QML leverages fundamental quantum properties such as superposition and entanglement, allowing for efficient processing of high-dimensional and complex data \cite{bib5,bib6,bib12}. Superposition enables quantum systems to explore multiple computational paths simultaneously, improving their ability to capture intricate dependencies within financial data. Entanglement facilitates stronger correlations between variables, leading to richer feature representations and enhanced pattern recognition that classical models struggle to achieve.
QML has been actively explored across various domains, including finance \cite{innan2024financial,dutta2024qadqn,innan2024financial,pathak2024resource}, healthcare \cite{rani2023quantum}, cybersecurity \cite{maouaki2024quantum}, natural language processing \cite{dave2024sentiqnf}, optimization \cite{innan2025optimizing}, and materials science \cite{innan2024quantum1,chen2024crossing}. These applications demonstrate the potential of quantum models in addressing computational challenges across diverse fields, further motivating research into their role in financial modeling.

Financial markets, in particular, present a complex and dynamic environment where QML can offer significant advantages. Challenges such as time-series forecasting, anomaly detection, high-frequency trading, risk management, and fraud detection require advanced analytical techniques for handling non-stationary and high-dimensional data. QML introduces quantum-enhanced feature spaces and optimization techniques, offering novel computational approaches for improving predictive analytics and decision-making in financial markets.

Extensive research has been conducted on the application of QML for financial modeling. Quantum-enhanced architectures, such as Quantum Neural Networks (QNNs) through Quantum Variational Circuits (QVCs) \cite{cerezo2021variational,abbas2021power}, have been proposed to complement or replace traditional deep learning approaches. These models employ quantum feature embedding, where data is transformed into a higher-dimensional quantum Hilbert space to improve separability and predictive accuracy \cite{bib7}. Additionally, hybrid quantum-classical models have emerged as a practical solution, bridging the gap between current quantum hardware limitations and real-world ML applications \cite{bib8}.
\begin{figure}[htpb]
    \centering
    \includegraphics[width=\linewidth]{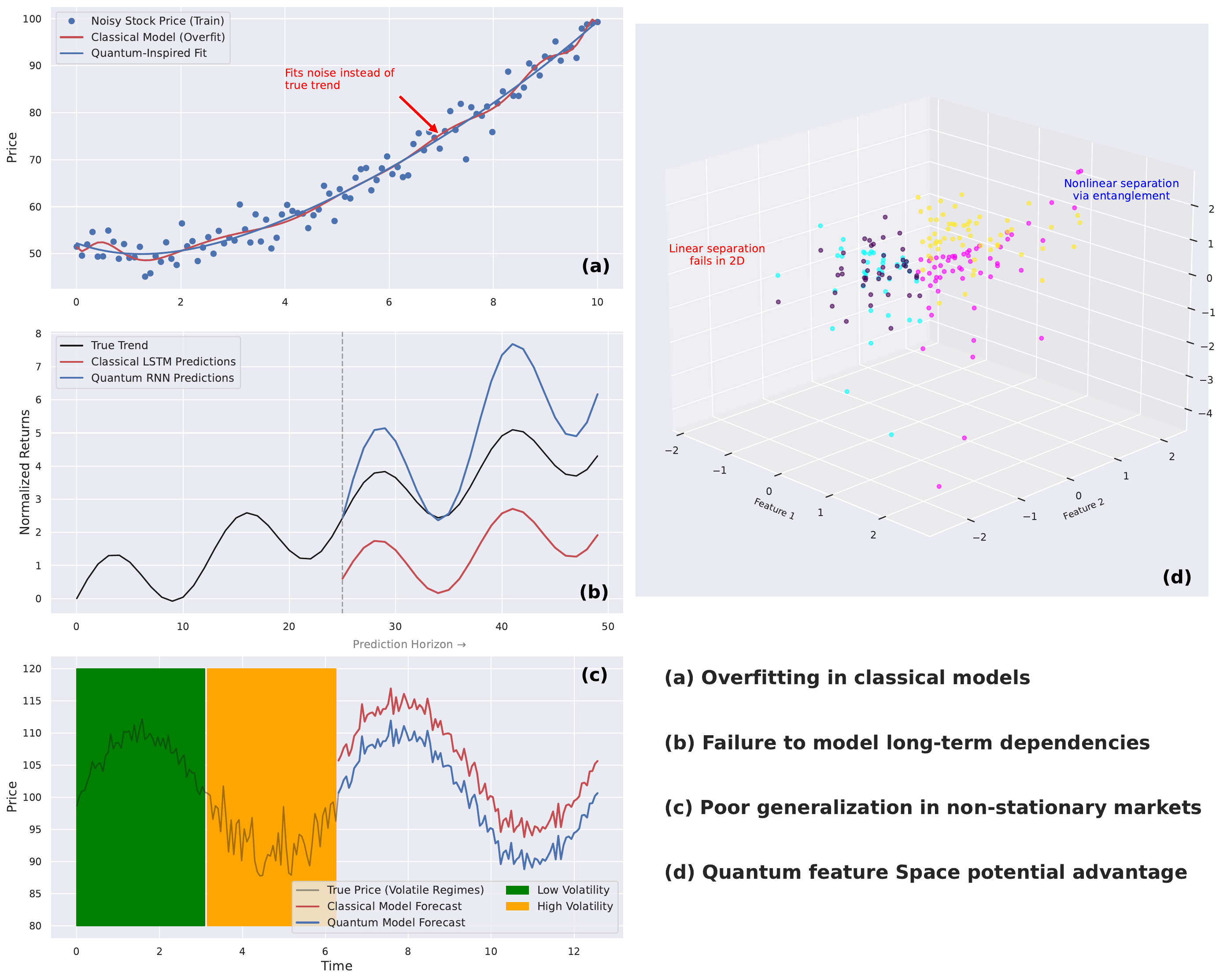}
    \vspace{-0.5cm}
    \caption{
    Schematic illustrating key challenges in classical modeling, motivating the exploration of quantum approaches. \textbf{(a)} Overfitting in classical models: A low-degree polynomial fit (pink) captures noise rather than the true trend, highlighting the limitations of classical curve-fitting techniques. \textbf{(b)} Failure to model long-term dependencies: While both classical LSTM (red) and quantum RNN (blue) models extend trends, neither fully captures complex dependencies, suggesting the need for improved architectures. \textbf{(c)} Poor generalization in non-stationary markets: Forecasts struggle in volatile regimes, where classical models show some alignment with the true trend, but quantum predictions exhibit larger deviations, raising questions about their robustness. \textbf{(d)} Quantum feature space: Nonlinear transformations in classical feature spaces can mimic certain quantum effects, yet the potential for quantum-enhanced representations remains an open question. Together, these findings highlight the need for further investigation into whether quantum models can provide genuine advantages beyond classical methods. \textit{The visualizations presented are conceptual representations rather than outputs from actual quantum models.}}
    \label{fig1}
\end{figure}

To evaluate the potential motivations for QML, we present a conceptual analysis contrasting classical and quantum methodologies using a mimic-based visualization approach. This method highlights fundamental limitations in classical financial models while identifying areas where quantum approaches may offer improvements. As illustrated in Fig.~\ref{fig1}, we categorize four key challenges in classical ML-based financial modeling:
\begin{itemize} 
\item \textbf{Overfitting in classical models:} Classical regression techniques, such as low-degree polynomial fits, often capture noise instead of the underlying trend, leading to poor generalization. This motivates the exploration of quantum models, which, leveraging higher-dimensional feature spaces, may enable more robust function approximation. 
\item \textbf{Failure to model long-term dependencies:} Classical recurrent architectures, including LSTMs, struggle to maintain long-range correlations. Although both classical and quantum RNNs attempt to extend temporal dependencies, the potential for quantum circuits to retain information more effectively over extended time horizons motivates their study. 
\item \textbf{Poor generalization in non-stationary markets:} Classical ML models often struggle in volatile regimes, exhibiting partial alignment with true trends but failing under abrupt shifts. The hypothesis that quantum approaches could enhance adaptability through their unique representational properties motivates further investigation. 
\item \textbf{Quantum feature space:} Classical models can mimic certain quantum effects via nonlinear transformations, but the possibility that quantum-enhanced representations may improve pattern recognition and decision-making motivates continued exploration of their role in financial forecasting. 
\end{itemize}

Despite its theoretical promise, the practical deployment of QML faces significant challenges. Quantum hardware remains constrained by noise, limited qubit reliability, and scalability issues, all impacting the feasibility of large-scale QML applications \cite{bib9,bib10}. The limited depth of current quantum circuits further restricts their ability to process high-dimensional financial data effectively \cite{bib11}. Given these constraints, hybrid quantum-classical models have emerged as a viable pathway for integrating QML into financial applications, employing QC for specific subproblems while relying on classical architectures for broader computations \cite{bib12}.

Recent research has explored the potential of hybrid quantum-classical models in financial forecasting and time-series prediction. Studies have demonstrated that Parameterized Quantum Circuits (PQCs), when integrated with classical optimization techniques, yield promising results in modeling complex financial patterns \cite{bib13,bib14}. In particular, hybrid models have shown superior capability in capturing nonlinear dependencies in financial data compared to standalone classical approaches \cite{bib15}. However, existing research has largely overlooked the incorporation of domain-specific technical indicators, such as the relative Strength Index (RSI), Moving Average Convergence Divergence (MACD), and Average Directional Index (ADX), which are widely used in financial analysis to enhance feature representations.

In this study, we introduce a novel hybrid quantum-classical framework for stock market prediction, integrating quantum variational circuits with classical deep learning architectures. Our approach enhances feature engineering by incorporating quantum-generated representations alongside conventional financial indicators, enriching the input space for predictive modeling. The key contributions of this research are as follows:
\begin{itemize}
\item \textbf{Development of a hybrid quantum-classical architecture:} We propose a hybrid model that seamlessly integrates custom-designed QNNs with classical deep learning architectures, including RNNs, LSTMs, BiLSTMs, and GRUs, to enhance predictive performance.
    \item \textbf{Introduction of a custom QNN regressor:} We design a QNN regressor based on a novel customized ansatz and parameterized quantum circuits, specifically optimized for financial time-series forecasting.
    \item \textbf{Dual hybrid model optimization strategies:} We explore two hybrid configurations: joint optimization of classical and quantum parameters and classical feature extraction followed by quantum modeling. Both approaches demonstrate measurable improvements in Root Mean Square Error (RMSE) compared to standalone QNNs.
    \item \textbf{Efficient parallelized model training and hyperparameter optimization:} We employ systematic hyperparameter tuning using TimeSeriesSplit, k-fold cross-validation, RMSE analysis, and predictive error visualization techniques to ensure robust model evaluation.
\end{itemize}
By systematically integrating QC techniques with classical deep learning methodologies, this research advances the state-of-the-art in financial time-series forecasting. Our findings provide valuable insights into the practical benefits and challenges of hybrid quantum-classical models, offering a new perspective on the role of QML in financial analytics.

The rest of the paper is organized as follows: Section \ref{sec2} provides an overview of the background and related work, outlining key advancements in QML for financial forecasting. Section \ref{sec3} details the proposed methodology, including the architecture of the proposed models. Section \ref{sec4} presents the results and discussion, analyzing the performance of the proposed architectures. Finally, Section \ref{sec5} summarizes our findings and highlights potential directions for future research.

\section{Related Work \label{sec2}}
\subsection{Financial Forecasting and ML}
Financial forecasting relies heavily on ML models to analyze historical data and predict future market trends. Traditional methods, such as Auto Regressive Integrated Moving Average (ARIMA) and Generalized Auto Regressive Conditional Heteroskedasticity (GARCH) \cite{bib30}, have been widely used to capture linear dependencies and volatility clustering in financial time series \cite{bib1}. Nonetheless, these techniques often struggle with the non-linearity and regime shifts observed in modern financial markets.
Deep learning approaches, particularly RNNs and LSTM networks, have addressed these challenges by capturing the complex temporal dependencies present in financial data \cite{bib26,bib2,bib23}. Recent work on advanced recurrent neural network architectures, including LSTM and GRU, demonstrated their effectiveness in modeling nonlinear and dynamic patterns, with subsequent studies supporting their role in improving forecasting performance \cite{bib27,bib37,bib38}.

Hybrid deep learning frameworks integrating attention mechanisms or external features, such as sentiment analysis, have also contributed to enhanced predictive performance \cite{bib19}. Existing work on RNN and LSTM architectures spans a wide range of applications and modeling techniques \cite{bib29}, while investigations into hybrid CNN-LSTM models indicate that merging convolutional neural networks for local feature extraction with LSTMs for sequential modeling can result in notable improvements in classification accuracy \cite{bib28}. Additional studies have incorporated alternative data sources, including social media sentiment, to enrich these hybrid frameworks \cite{bib39,bib40}.

Past investigations have demonstrated that such encoding can boost ML model performance by aiding in the identification of momentum shifts and trend reversals \cite{bib19,bib20,bib31}. Further work on algorithmic trading and financial time series employing deep convolutional neural networks emphasizes the growing importance of advanced feature engineering, with additional studies exploring the benefits of integrating macroeconomic variables and alternative data to further enhance forecasting accuracy \cite{bib33,bib32,bib41,bib42}.
Despite these advancements, classical ML models encounter challenges, including overfitting, difficulty capturing long-range dependencies, and limited adaptability in non-stationary markets \cite{bib3,bib4}.

\subsection{Financial Forecasting and QML}
Recent advances in quantum-enhanced approaches have sought to address the complexity and volatility inherent in modern financial markets. For example, embedding data in quantum circuits has improved feature separability \cite{bib7}, thereby enhancing trend detection in financial time series. In addition, hybrid quantum-classical models that integrate quantum-enhanced LSTM networks have been applied to stock price prediction \cite{bib34}. Prior work on QML in finance has explored its role in time series forecasting \cite{bib36, bib37} and its broader implications for financial modeling \cite{bib35}. Further strategies have been proposed to address the challenges of adopting QML in finance, with novel designs aimed at improving financial analysis and decision-making \cite{bib38, bib39}.

Beyond these developments, quantum-enhanced reinforcement learning has been explored for financial trading. One approach employed a quantum attention deep Q-network combining a variational quantum circuit with a deep Q-learning framework, achieving superior risk-adjusted returns on major market indices under realistic conditions \cite{dutta2024qadqn}. Another hybrid design uses an encoder to convert financial time series into density matrices and a deep quantum network to predict future matrices, with a subsequent classical network inferring price levels; empirical evaluations on 24 securities demonstrate both accuracy and efficiency \cite{paquet2022quantumleap}. Quantum methods have also been applied to high-frequency trading, where processing large volumes of rapidly arriving data is essential \cite{palaniappan2024review}, and quantum neural networks have been investigated for stock price prediction and financial engineering to help mitigate losses and guide trading decisions \cite{iyer2024artificial}.

Interpretability remains a critical requirement in financial forecasting, as decisions must be supported by transparent reasoning. While Shapley Additive exPlanations (SHAP) have been widely used to attribute feature contributions in classical models \cite{bib22}, their application in QML remains underexplored. Recent investigations into quantum representation learning combined with explainable artificial intelligence \cite{bib41} and studies on feature importance in quantum frameworks \cite{bib40} underscore the need for more interpretable quantum-enhanced models. In this context, our work incorporates these interpretability techniques to clarify the decision-making process within hybrid quantum-classical models.

\section{Methodology \label{sec3}}
\begin{figure*}
    \centering    
    \includegraphics[width=1\linewidth]{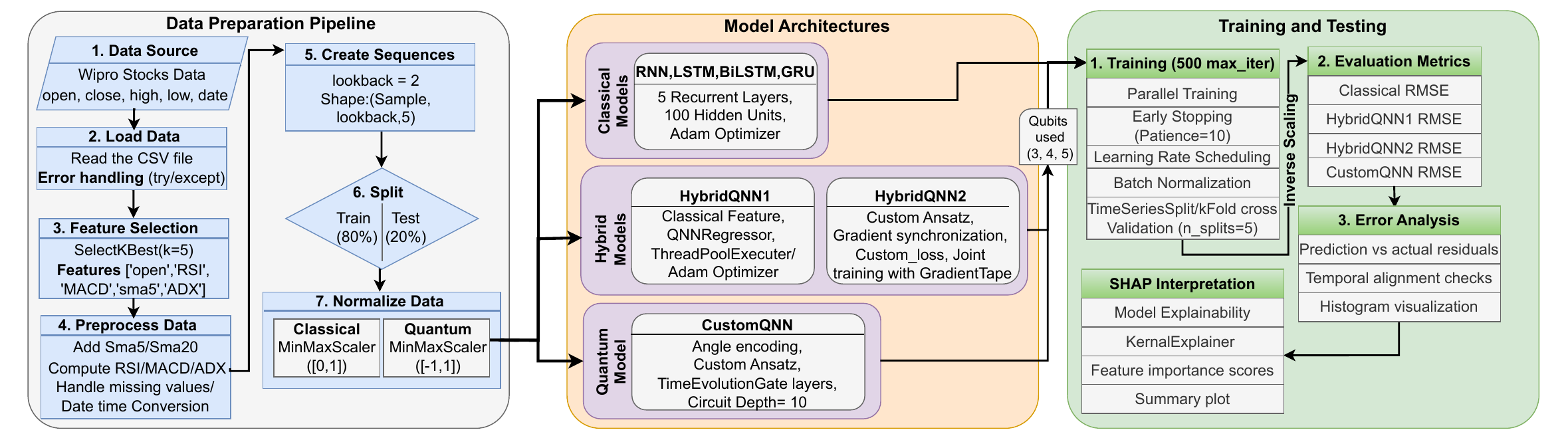}
    \vspace{-0.6cm}
    \caption{Overview of the proposed hybrid quantum-classical methodology.}
    \label{fig:methodology}
\end{figure*}

This study presents the design and implementation of a hybrid quantum-classical model for stock market prediction, integrating classical ML with quantum circuits. The methodology includes data preparation, model architectures, and evaluation metrics. A comprehensive overview of the approach is illustrated in Fig.~\ref{fig:methodology}.

\subsection{Data Preparation and Preprocessing}
The dataset employed in this study comprises historical stock prices, specifically the open, close, high, and low values. To capture a more comprehensive representation of market behavior, the dataset is augmented with several technical indicators. In particular, the RSI is computed over a 14-day period to capture market momentum and identify potential overbought or oversold conditions. Additionally, the MACD is calculated by combining short-term (12-day) and long-term (26-day) exponential moving averages, complemented by a 9-day signal line, providing insights into trend reversals and shifts in market momentum. Finally, the ADX is determined over a 14-day period to assess the strength of prevailing trends. The integration of these technical indicators enriches the dataset, facilitating a more robust analysis of market dynamics.

To preserve temporal integrity, the dataset is split in a time-ordered manner into training and testing subsets. Sequential data structures are then constructed based on a predefined lookback period, ensuring that the model learns meaningful patterns from past stock price movements. However, using all available features may introduce noise and redundancy, which can degrade model performance. Therefore, feature selection is applied to refine the dataset by identifying the most relevant inputs.

To achieve this, the SelectKBest method is used to extract the most informative features that best describe the target variable, which, in this case, is the closing price. These selected features are then structured into input sequences aligned with the predefined lookback period, ensuring the model captures the essential temporal dependencies within the stock data. Since models are sensitive to feature scale variations, an additional normalization step is necessary to maintain numerical stability and facilitate efficient learning.

To ensure consistent feature magnitudes and promote stable model convergence, data normalization techniques are applied. Specifically, the feature data are scaled to a range between 0 and 1 using a MinMaxScaler, thereby ensuring uniformity across all input features. Similarly, the target variable is normalized to facilitate efficient learning in both the classical and quantum components of the model.

\subsection{Model Architectures}

The proposed methodology integrates both classical and quantum ML models to predict stock market trends. Classical models, including RNN, LSTM, Bidirectional LSTM (BiLSTM), and Gated Recurrent Units (GRU), serve as benchmark models to evaluate the effectiveness of hybrid quantum-classical learning. These deep learning architectures are widely used for time-series forecasting due to their ability to capture temporal dependencies. However, our primary focus is on quantum-enhanced methods, exploring their potential to uncover complex correlations that classical models may not efficiently capture.

To achieve this, we propose three quantum-driven approaches: a standalone QNN, which is developed from a Hamiltonian-based PQC, and two hybrid models that integrate classical deep learning with quantum computation. The first hybrid approach, HybridQNN1, applies a classical deep learning network for initial feature extraction before passing the transformed data to a quantum circuit for further processing. The second approach, HybridQNN2, employs an end-to-end quantum-classical framework, where the classical and quantum components are jointly optimized, ensuring seamless interaction between the two paradigms.

\subsubsection{Customized QNN}

The QNN is the core computational framework, leveraging a PQC inspired by a Hamiltonian formulation. This design enables the quantum model to exploit entanglement and quantum interference, capturing intricate dependencies in financial data.

Initially, all qubits are prepared in the ground state:

\begin{equation}
|\psi(0)\rangle = |0\rangle^{\otimes n}.
\end{equation}
\begin{figure}[htpb]
    \centering
    
    \includegraphics[width=0.45\linewidth]{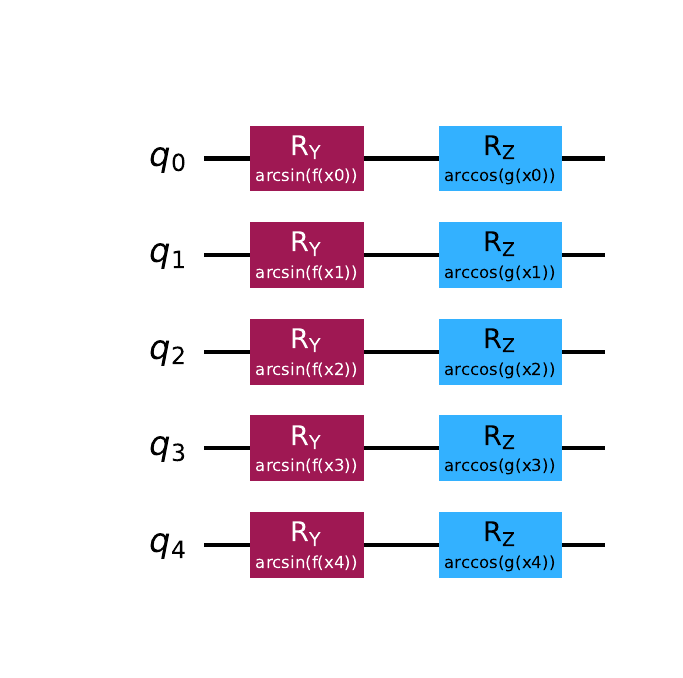}
    \caption{Angle encoding circuit, where each qubit \(q_i\) is encoded using a parameterized \(R_Y\) gate with angle \(\arcsin(f(x_i))\) followed by an \(R_Z\) gate with angle \(\arccos(g(x_i))\), mapping the classical input \(x_i\) into a quantum state via nonlinear transformations.}

    \label{fig:encoding}
\end{figure}
Classical stock market data is mapped onto quantum states using angle encoding (see Fig. \ref{fig:encoding}). Each feature \( x_i \) is embedded via single-qubit rotation gates:

\begin{equation}
R_Y(\theta_i) = \exp\left(-i \frac{\theta_i}{2} Y\right), \quad
R_Z(\phi_i) = \exp\left(-i \frac{\phi_i}{2} Z\right),
\end{equation}
where \( \theta_i = \arcsin(f(x_i)) \) and \( \phi_i = \arccos(g(x_i)) \), ensuring an efficient transformation of classical information into the quantum domain. The Hamiltonian corresponding to these gates are given by:

\begin{equation}
    H_{\text{single}} = \sum_{i} \left(\frac{\theta_i}{2} \sigma_y^{(i)} + \frac{\phi_i}{2} \sigma_z^{(i)}\right).
\end{equation}

To enhance expressivity and mitigate training issues such as barren plateaus, we design a customized ansatz composed of parameterized single-qubit rotations and entangling operations. The entanglement structure, introduced through controlled-NOT (CNOT) and controlled-Z (CZ) gates, establishes quantum correlations between qubits, allowing the model to capture non-trivial relationships in stock market trends.

The Hamiltonians governing these entangling operations are given by:

\begin{equation}
H_{\text{CNOT}} = \sum_{\langle i, j \rangle} \frac{1}{2} \left(1 - \sigma_z^{(i)}\right) \sigma_x^{(j)},
\end{equation}
where $\sigma_z^{(i)}$ is the Pauli-Z operator acting on qubit $i$, determining whether the control qubit is in state $\lvert 1 \rangle$, and $\sigma_x^{(j)}$ is the Pauli-X (bit-flip) operation applied to qubit $j$ if qubit $i$ is in $\lvert 1 \rangle$.

\begin{equation}
H_{\text{CZ}} = \sum_{\langle i, j \rangle} \frac{1}{4} \left(1 - \sigma_z^{(i)}\right) \left(1 - \sigma_z^{(j)}\right).
\end{equation}

The overall Hamiltonian governing the QNN is:

\begin{equation}
H = H_{\text{single}} + H_{\text{CNOT}} + H_{\text{CZ}},
\end{equation}
where \( H_{\text{single}} \) encapsulates all parameterized single-qubit rotations across multiple layers. The circuit architecture is designed to optimize the balance between expressivity and trainability, ensuring robust learning performance.

Once quantum transformations are applied, measurements are performed in the computational basis to extract relevant statistical features for prediction. The probability of obtaining a measurement outcome \( x \) is given by:

\begin{equation}
P(x) = |\langle x | \psi(t)\rangle|^2.
\end{equation}

This probability distribution is then used to compute observables relevant to financial modeling.

\begin{figure}[htpb]
    \centering
    \includegraphics[width=0.5\textwidth]{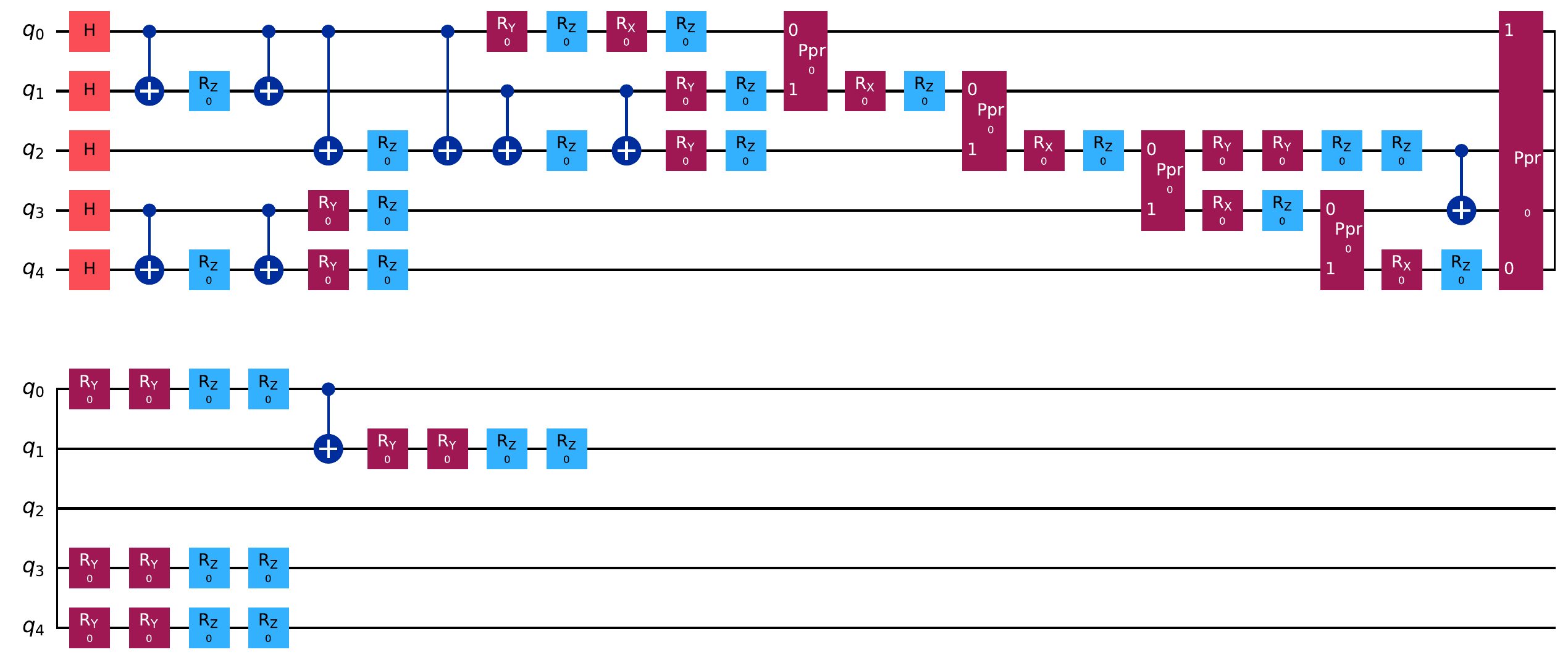}
    \vspace{-0.6cm}
    \caption{Customized quantum ansatz utilized in QNN (also in HybridQNN2). The circuit integrates parameterized single-qubit rotations and entangling operations to enhance feature representation and learning capacity.}
    \label{fig:Ansatz-circuit}
\end{figure}

The customized quantum ansatz utilized in our QNN framework is depicted in Fig.~\ref{fig:Ansatz-circuit}, illustrating its layered structure with entangling operations and parameterized quantum gates.

\subsubsection{HybridQNN1: Classical Preprocessing with Quantum Processing}
The HybridQNN1 model follows a sequential classical-to-quantum pipeline, where a classical deep learning model first extracts meaningful features from stock market data before passing the transformed features to a shallow quantum circuit for further refinement. This approach ensures that the classical model efficiently captures temporal dependencies, while the quantum circuit enhances feature representation through entanglement and interference. The process starts with \( X \), representing the input stock price data, which undergoes feature extraction through a classical model before being encoded into quantum states. The quantum model processes the encoded data, applying variational transformations through a QNN regressor~\cite{bib7}. Finally, the measured quantum outputs are passed to a post-processing layer before generating the final stock price prediction. 

The classical feature extraction module is implemented by combining both handcrafted feature selection using SelectKBest with regression and deep learning based feature extraction via an LSTM network~\cite{bib42,bib2}, which is well-suited for capturing long-range dependencies in time-series data. Given an input sequence of stock prices \( X = \{x_1, x_2, ..., x_t\} \), the classical model computes: 
\begin{equation} 
h_t = \sigma(W_h h_{t-1} + W_x x_t + b_h), 
\end{equation} 
\begin{equation} 
o_t = W_o h_t + b_o, 
\end{equation} 
where \( h_t \) represents the hidden state at time step \( t \), \( W_h, W_x, W_o \) are learnable weight matrices, \( b_h, b_o \) are biases, and \( \sigma \) is the activation function. The final hidden state of the classical model serves as the feature representation, which is subsequently encoded into a quantum state for further quantum processing.  

\begin{figure}[htpb] 
\centering 
\includegraphics[width=\linewidth]{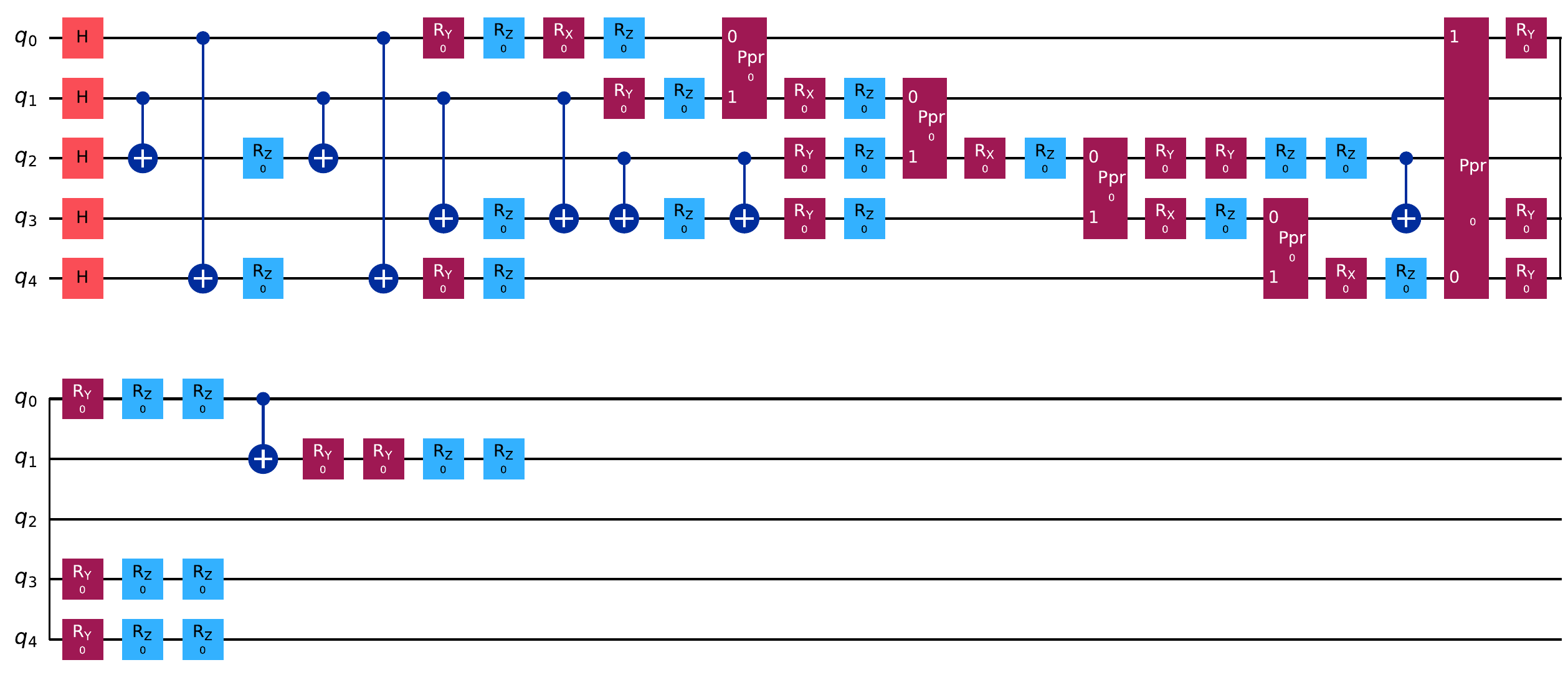} 
\vspace{-0.6cm} 
\caption{QNN Regressor circuit, the core quantum processing unit in HybridQNN1. The circuit integrates parameterized single-qubit operations and entangling gates to enhance feature learning.} 
\label{fig:HybridQNN1} 
\end{figure}

Following classical feature extraction, the processed data is encoded into quantum states using the same angle encoding method as explained in the QNN. Once the quantum state is prepared, the RegressorQNN applies a sequence of variational transformations, consisting of trainable single-qubit rotations and entangling operations (CNOT, CZ). These transformations introduce non-classical correlations, enabling the quantum circuit to capture higher-order dependencies in financial data. Following quantum processing, measurements are performed on a computational basis, similar to the QNN. The architecture of the QNN Regressor, a key component of HybridQNN1, is illustrated in Fig.~\ref{fig:HybridQNN1}.

\subsubsection{HybridQNN2: Fully Integrated Quantum-Classical Optimization}

Unlike HybridQNN1, which applies quantum processing after classical feature extraction, in HybridQNN2, both the classical and quantum components are trained in a fully integrated manner through joint optimization. This design ensures that raw stock price features are processed concurrently by a classical LSTM-based feature extractor and a QNN based on a Hamiltonian-derived PQC. The classical network captures broader market trends, while the QNN is designed to uncover intricate correlations that may be less accessible via classical methods.

\begin{figure*}[htpb]
    \centering
    \includegraphics[width=\textwidth]{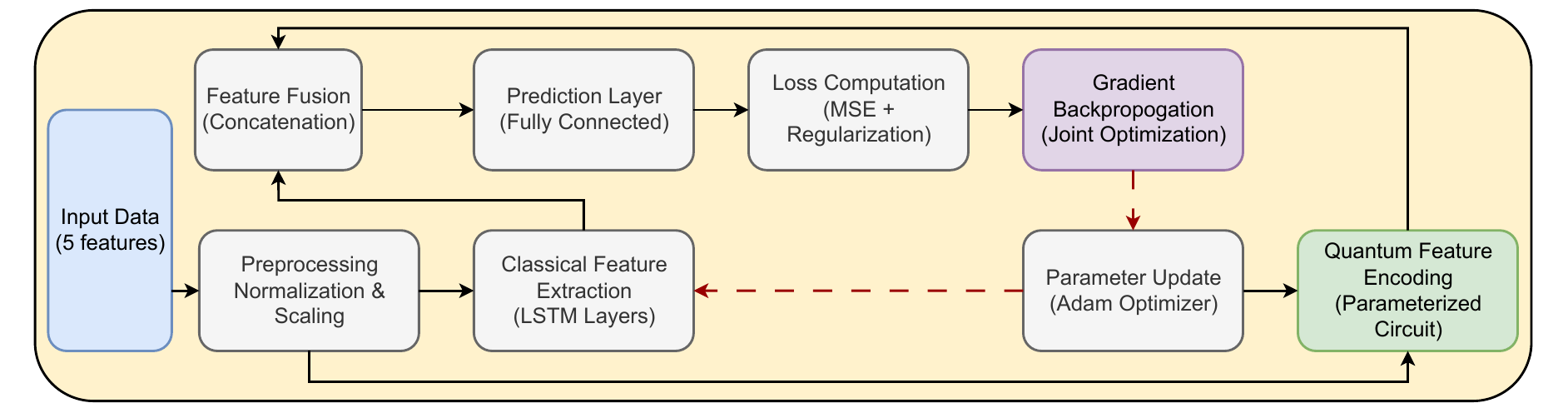}
    \vspace{-0.5cm}
    \caption{HybridQNN2 architecture, the process starts with the preprocessed data being simultaneously fed to a classical LSTM-based feature extractor and a parameterized quantum circuit. Their outputs are fused and passed to a fully connected prediction layer, with joint backpropagation updating both components.}
    \label{fig:hybrid_joint_integration}
\end{figure*}

As shown in Fig.~\ref{fig:hybrid_joint_integration}, preprocessed data is fed in parallel to both the classical and quantum layers. Their outputs are fused in a dedicated fusion layer before being passed on to a fully connected prediction layer. This entire process is optimized end-to-end using backpropagation with shared loss functions, enabling cohesive learning across both components.

Mathematically, the fusion process can be expressed as:
\begin{equation}
X_{\text{hybrid}} = f(W_{\text{quantum}} \cdot X_{\text{QNN}} + W_{\text{classical}} \cdot X_{\text{DL}}),
\end{equation}
where \(W_{\text{quantum}}\) and \(W_{\text{classical}}\) are trainable parameters, \(X_{\text{QNN}}\) and \(X_{\text{DL}}\) represent the quantum and classical feature outputs respectively, and \(f(\cdot)\) is a nonlinear activation function.

This fully integrated quantum-classical model not only simplifies the training process but also leverages the strengths of both components. The simultaneous optimization enhances generalization performance, offering robust predictive capabilities by combining quantum-enhanced feature extraction with the established power of deep learning.

\subsection{Training Process and Optimization}

For all three models—customized QNN, HybridQNN1, and HybridQNN2—the training process is designed to optimize the entire network by updating all trainable parameters, including those within the quantum circuit, using a classical optimizer. In our approach, the classical ADAM optimizer is employed to update the parameters in a unified manner, ensuring that both the quantum and classical components are refined concurrently via backpropagation.

The training procedure begins with the partitioning of the time-series data using TimeSeriesSplit cross-validation. This method maintains the temporal order of the dataset, enabling an authentic evaluation of the model's performance across sequential data folds. To further enhance generalization and prevent overfitting, early stopping is implemented by monitoring the validation loss, interrupting the optimization process when improvements become negligible. Additionally, a learning rate schedule is applied to gradually reduce the learning rate during training, which facilitates more precise convergence.

The core of the optimization strategy lies in the joint update of all trainable parameters—both those in the classical layers and those within the quantum circuit. The ADAM optimizer, a robust variant of stochastic gradient descent, is used to minimize the Mean Squared Error (MSE) cost function defined as:
\begin{equation}
\text{Cost} = \frac{1}{n} \sum_{i=1}^n (y_i - \hat{y}_i)^2,
\end{equation}
where \(y_i\) represents the true target values, \(\hat{y}_i\) are the predicted values, and \(n\) is the number of samples. During each iteration, the optimizer computes gradients with respect to the loss and updates the corresponding parameters in both the classical deep learning modules and the quantum circuit. This unified update mechanism ensures that improvements in one component are propagated throughout the network, leading to a more cohesive learning process.

This training process integrates classical optimization techniques with joint parameter updates to optimize both the quantum and classical components. By employing TimeSeriesSplit cross-validation, early stopping, and a learning rate schedule alongside the ADAM optimizer, our framework effectively refines all trainable parameters, resulting in enhanced performance and robust generalization across customized QNN, HybridQNN1, and HybridQNN2 models.

\section{Results and Discussion \label{sec4}}
\subsection{Experimental Settings}
To ensure a rigorous evaluation of the proposed hybrid quantum-classical models for stock market prediction, the experimental setup incorporated well-defined dataset preparation, model architectures, hyperparameter tuning, and evaluation metrics. The details are summarized in Table~\ref{tab:experimental_settings}.
\begin{table}[htpb] 
\centering 
\caption{Summary of our experimental settings.} \label{tab:experimental_settings} 
\begin{adjustbox}{width={\linewidth}}
\begin{tabular}{lc} 
\toprule 
\textbf{Parameter} & \textbf{Details} \\ 
\midrule 
\textbf{Dataset} & Historical stock prices~\cite{bib43} \\ 
\textbf{Data Partitioning} & 80\% training, 20\% testing \\ \textbf{Validation Methods} & TimeSeriesSplit and k-fold cross-validation \\ 
\textbf{Classical Models} & LSTM, RNN, BiLSTM, GRU \cite{bib3,bib4,bib18} \\ 
\textbf{Quantum Frameworks} & Qiskit and scikit-QULACS \cite{javadi2024quantum,scikit-QULACS} \\ 
\textbf{Classical Frameworks} & TensorFlow and scikit-learn \\ 
\textbf{Number of Qubits} & 3, 4, and 5 \\ 
\textbf{Hyperparameters} & Lookback period: 2, Batch size: 32, Max iterations: 500, Early stopping applied \\ \textbf{Optimization Algorithm} & ADAM optimizer with learning rate scheduling \\ 
\textbf{Performance Metric} & RMSE \\ \bottomrule 
\end{tabular} 
\end{adjustbox}
\end{table}

Experiments are conducted using a quantum simulator (QULACS) and a High-Performance Computing (HPC) cluster. The classical and hybrid quantum-classical models were trained on the PARAM Shivay supercomputer at IIT (BHU), utilizing GPU nodes equipped with 2 × Intel Xeon Skylake 6148 CPUs (20 cores @ 2.4GHz each), 192GB RAM, and 2 × NVIDIA Tesla V100 GPUs (5120 CUDA cores, 16GB HBM2). The code execution is parallelized using CUDA and scheduled via Slurm job management, leveraging 100Gbps InfiniBand EDR network communication for efficient multi-node execution.
\subsection{Stock Price Prediction Performance}
\begin{figure}[htpb]
    \centering
    \includegraphics[width=\linewidth]{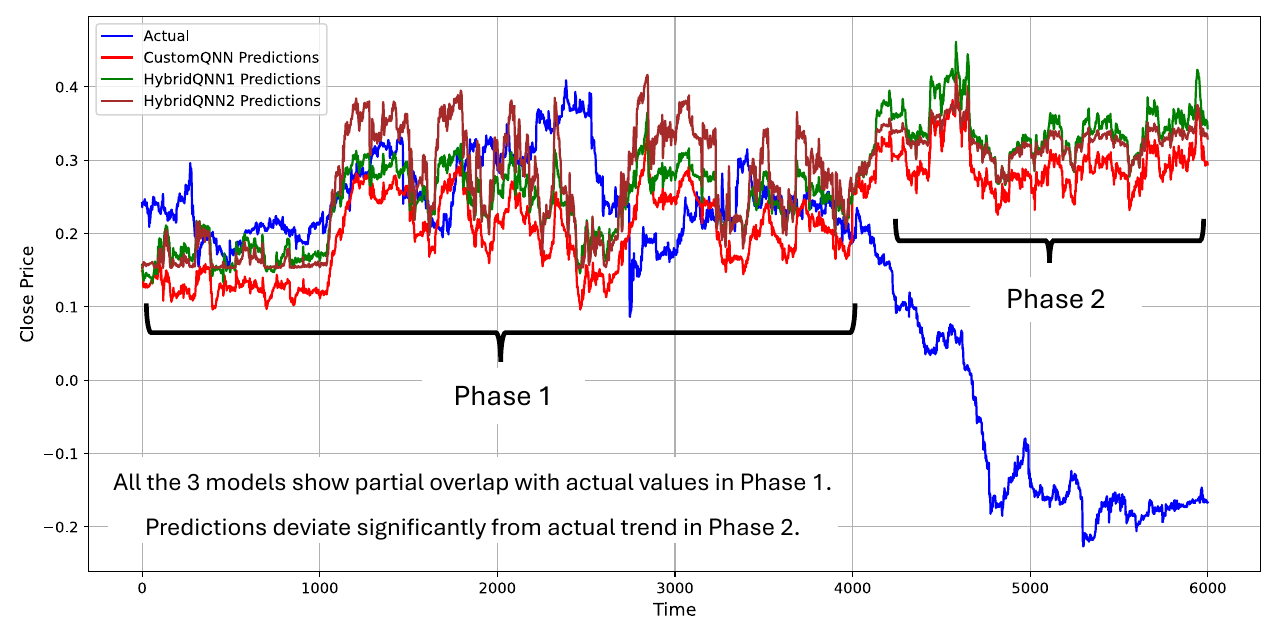}
    \vspace{-0.7cm}
    \caption{Comparison of actual stock prices with predictions from quantum and hybrid QNN models: CustomQNN, HybridQNN1, and HybridQNN2. Phase 1 (Time: 0–4000) demonstrates a reasonable overlap between predictions and actual values, while in Phase 2 (Time: 4000–6000), predictions fail to capture the sharp downward trend in stock prices.}
    \label{fig:predictions}
\end{figure}

 We assess the predictive performance of quantum and hybrid QNN models in forecasting stock price trends.
As shown in Fig.~\ref{fig:predictions}, in Phase 1 (Time: 0–4000), the actual stock prices fluctuate within a relatively narrow range (Close Price: 0.25–0.45). The predictions from CustomQNN, HybridQNN1, and HybridQNN2 exhibit partial overlap with these real values, effectively capturing general price levels and broader trends, albeit with occasional deviations in short-term fluctuations.
In Phase 2 (Time: 4000–6000), the actual stock prices experience a sharp downward shift toward values close to zero. However, the predictions from all three models remain consistently within the initial price range (0.25–0.45), indicating their inability to promptly adjust to sudden market changes or regime shifts. This discrepancy underscores a limitation in model responsiveness to rapid price dynamics.
Overall, while the evaluated models demonstrate promising predictive capability during stable market conditions (Phase 1), their performance deteriorates significantly during abrupt market transitions (Phase 2). This highlights potential areas for improvement in quantum and hybrid models through further optimization and adaptive training strategies.

\subsection{Prediction Error Analysis}

\begin{figure}[htpb]
     \centering
     \includegraphics[width=\linewidth]{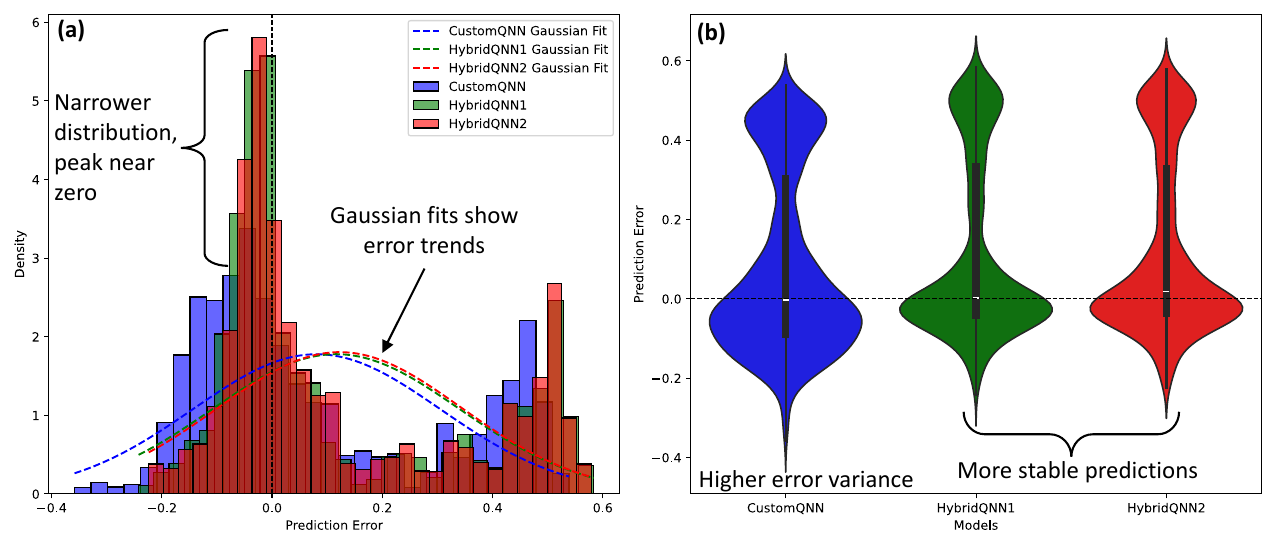}
     \vspace{-0.6cm}
     \caption{Comparison of prediction error distributions across CustomQNN, HybridQNN1, and HybridQNN2. \textbf{(a)} The histogram of prediction errors with Gaussian fits shows that HybridQNN models exhibit a narrower distribution with a peak near zero, indicating lower variance and more consistent predictions. In contrast, CustomQNN displays a wider spread, suggesting higher variability in errors.
\textbf{(b)} The violin plots further highlight the stability of HybridQNN models, showing a tighter error distribution compared to CustomQNN, which demonstrates greater variance.}
     \label{fig:error_distribution}
\end{figure}

To further evaluate the stock price prediction performance, we analyze the models' error distributions using Gaussian fit analysis and violin plot interpretation. As shown in Fig. \ref{fig:error_distribution}-a, the histogram provides a detailed view of the error spread, while the Gaussian fit illustrates how well the errors follow a normal distribution. The CustomQNN model exhibits a broader distribution, indicating a higher variance in its prediction errors. In contrast, HybridQNN1 and HybridQNN2 demonstrate narrower distributions, suggesting that their errors are more concentrated around zero. The tighter Gaussian fits of HybridQNN models confirm their greater stability and lower error variance, though they are still not perfect.

In Fig. \ref{fig:error_distribution}-b, the violin plot provides additional insight by visualizing the density and spread of errors for each model. The CustomQNN model displays a wider shape with notable outliers, reflecting its higher variability. On the other hand, HybridQNN1 and HybridQNN2 show more compact distributions, with errors clustered near zero, indicating improved consistency and reduced variance in their predictions.

\subsection{Qubit Count Analysis}

To assess the impact of qubit count on model performance, we analyze the loss values and training times across different qubit configurations for CustomQNN, HybridQNN1, and HybridQNN2. As shown in Table \ref{tab:qubit_analysis}, we compare these models using 3, 4, and 5 qubits, highlighting their respective RMSE values and computational costs. CustomQNN consistently exhibits higher RMSE values, indicating less accurate predictions, whereas HybridQNN1 and HybridQNN2 achieve lower RMSE values, with HybridQNN2 maintaining the lowest error across all configurations. The 5-qubit setting yields the best accuracy, particularly for HybridQNN2, reinforcing the benefit of more tightly integrated quantum-classical layers. However, computational trade-offs emerge as training time increases with higher qubit counts. While HybridQNN2 is more computationally efficient in lower qubit configurations, HybridQNN1 exhibits a sharp rise in training time at 5 qubits. This highlights a key trade-off between accuracy and efficiency, where increasing qubit counts can enhance predictive performance and introduce higher computational costs.

 \begin{table}[htpb]
     \centering
     \caption{RMSE and training time (in seconds) for CustomQNN, HybridQNN1, and HybridQNN2 across different qubit configurations using TimeSeriesSplit. The results illustrate the trade-off between prediction accuracy and computational cost, with HybridQNN2 achieving the lowest RMSE while maintaining relatively efficient training times.}
     \label{tab:qubit_analysis}
     \begin{adjustbox}{width={\linewidth}}
     \begin{tabular}{lcccccc} 
         \toprule
         \textbf{Model} & \multicolumn{3}{c}{\textbf{Average RMSE}} & \multicolumn{3}{c}{\textbf{Training Time (Seconds)}} \\
         \cmidrule(lr){2-4} \cmidrule(lr){5-7}
          & \textbf{3 Qubits} & \textbf{4 Qubits} & \textbf{5 Qubits} & \textbf{3 Qubits} & \textbf{4 Qubits} & \textbf{5 Qubits} \\
         \midrule
         CustomQNN  & 0.07603 & 0.05528 & 0.06120 & 120765.63  & 155362.05  & 139337.06   \\
         HybridQNN1 & 0.02605 & 0.02161 & 0.01740 & 121020.84  & 155336.05  & 227781.18  \\
         HybridQNN2 & 0.02312 & 0.01959 & 0.01920 & 69841.69  & 92337.84  & 118833.32   \\
         \bottomrule
     \end{tabular}
     \end{adjustbox}
 \end{table}

\subsection{Loss Analysis Using k-Fold Cross-Validation and Time Series Techniques}

We validate model robustness by analyzing loss values under two distinct cross-validation strategies: k-Fold Cross-Validation and TimeSeriesSplit. The k-Fold Cross-Validation technique randomly divides the dataset into K equal folds, assuming data points are independently and identically distributed. While effective for many datasets, this approach can introduce information leakage in time-series data, as future observations may unintentionally appear in the training set. TimeSeriesSplit, in contrast, maintains the chronological order of data, ensuring that training is based only on past data while testing is conducted on future observations, making it more suitable for sequential dependencies.

\begin{figure}[htpb]
     \centering
     \includegraphics[width=\linewidth]{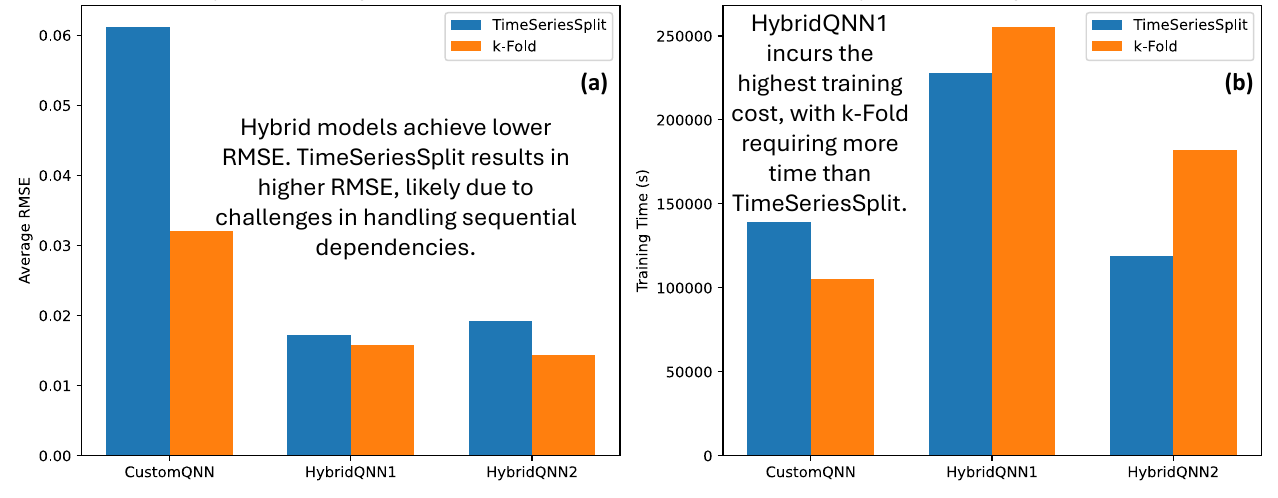}
     \vspace{-0.5cm}
     \caption{Comparison of RMSE and training time for our models using TimeSeriesSplit and k-Fold cross-validation.\textbf{ (a) }RMSE decreases across models, indicating improved predictive accuracy, while TimeSeriesSplit results in higher RMSE due to challenges in handling sequential dependencies. \textbf{(b)} Training time varies significantly across models, with HybridQNN1 incurring the highest computational cost, and k-Fold generally requiring more time than TimeSeriesSplit. The results suggest a trade-off between predictive accuracy and computational efficiency in hybrid quantum models.}
     \label{fig:Timeserieskfold}
 \end{figure}

As shown in Fig.~\ref{fig:Timeserieskfold}-a, RMSE consistently decreases across models, suggesting that k-Fold cross-validation provides a more generalized estimate of predictive performance, while TimeSeriesSplit results in higher RMSE due to the inherent challenges of handling sequential dependencies. Among the models, HybridQNN2 consistently achieves the lowest RMSE, reinforcing its superior predictive accuracy. However, this improvement comes at a computational cost. Fig.~\ref{fig:Timeserieskfold}-b highlights the training time variations across models, where HybridQNN1 incurs the highest computational cost, particularly when using k-Fold cross-validation, which generally requires more training time than TimeSeriesSplit due to repeated model evaluations across multiple folds.

These results suggest a trade-off between predictive accuracy and computational efficiency when selecting a validation approach for hybrid quantum models. While k-Fold cross-validation provides a more stable performance estimate, TimeSeriesSplit aligns better with real-world sequential forecasting tasks. The choice between these methods depends on the application requirements, balancing generalization, computational cost, and model reliability in dynamic financial environments.

\subsection{Comparative Performance Analysis}
\begin{figure}[htpb]
    \centering
    \includegraphics[width=\linewidth]{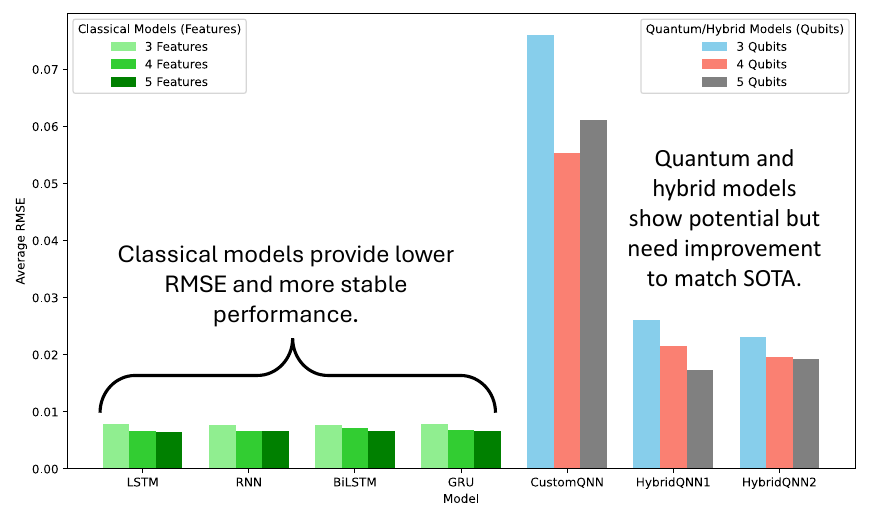}
    \vspace{-0.8cm}
    \caption{Comparison of average RMSE for classical models (LSTM, RNN, BiLSTM, GRU) with 3, 4, and 5 selected features (via SelectKBest) and our quantum/hybrid models (CustomQNN, HybridQNN1, HybridQNN2) with 3, 4, and 5 qubits, respectively. The number of selected features in classical models corresponds to the number of qubits in quantum/hybrid models. Classical models exhibit stable RMSE with limited improvement from additional features, while quantum models show potential but require further optimization. Hybrid models demonstrate better performance than pure quantum models.}
    \label{fig:Average_RMSE_Comparison}
\end{figure}

Rather than focusing solely on comparing quantum and classical models, this section evaluates the effectiveness of hybrid quantum-classical architectures within the broader landscape of stock price prediction. As shown in Table~\ref{tab:comparison_results} and Fig.~\ref{fig:Average_RMSE_Comparison}, HybridQNN models consistently achieve lower RMSE than CustomQNN, demonstrating the benefits of hybridization. However, classical models continue to achieve the lowest RMSE, reinforcing their robustness and efficiency in this task.

A clear trend emerges: classical models maintain stable RMSE values with minimal improvement from additional features, whereas hybrid models exhibit stronger gains in predictive performance compared to purely quantum models. Notably, HybridQNN2 achieves the lowest RMSE among quantum-enhanced methods, confirming that hybridization mitigates some quantum limitations for this case. Despite these advantages, hybrid models still face computational trade-offs. CustomQNN experiences high RMSE and prolonged training times, highlighting inefficiencies in standalone quantum models. HybridQNN1, while performing well, incurs significantly higher training costs than HybridQNN2, suggesting that different quantum-classical integration strategies influence overall efficiency.

These findings emphasize the promise of hybrid architectures while highlighting areas for further improvement. Optimizing quantum encoding schemes, reducing circuit depth, and enhancing noise resilience could enhance the feasibility of hybrid approaches. As quantum hardware continues to advance, hybrid models may become more competitive with classical methods, narrowing the performance gap and offering new possibilities for financial forecasting and other predictive tasks.

 \begin{table}[htpb]
     \renewcommand{\arraystretch}{1.2}
    \setlength{\tabcolsep}{4pt}   
 
     \centering
     \caption{
     Comparison of Average RMSE and Training Time (seconds) for models using 3, 4, and 5 selected features, determined via \texttt{SelectKBest} (k=3, k=4, k=5). The selected feature count is mapped to the number of qubits in quantum models. Results are obtained using TimeSeriesSplit cross-validation.}
     \label{tab:comparison_results}
\begin{adjustbox}{width={\linewidth}}

     \begin{tabular}{lcccccc}
         \toprule
         \textbf{Model} & \multicolumn{3}{c}{\textbf{Average RMSE}} & \multicolumn{3}{c}{\textbf{Training Time (Seconds)}} \\
         \cmidrule(lr){2-4} \cmidrule(lr){5-7}
          & \textbf{3 Features} & \textbf{4 Features} & \textbf{5 Features} & \textbf{3 Features} & \textbf{4 Features} & \textbf{5 Features} \\
         \midrule
         LSTM       & 0.00781 & 0.00670 & 0.00649 & 3790.59 & 3281.62 & 2735.92  \\
         RNN        & 0.00772 & 0.00671 & 0.00659 & 3699.21 & 3139.98 & 2232.31  \\
         BiLSTM     & 0.00775 & 0.00715 & 0.00669 & 4146.64 & 3675.25 & 3471.38  \\
         GRU        & 0.00781 & 0.00687 & 0.00669 & 3806.17 & 3319.52 & 2854.31  \\
         CustomQNN  & 0.07603 & 0.05528 & 0.06120 & 120765.63  & 155362.05  & 139337.06   \\
         HybridQNN1 & 0.02605 & 0.02161 & 0.01740 & 121020.84  & 155336.05  & 227781.18  \\
         HybridQNN2 & 0.02312 & 0.01959 & 0.01920 & 69841.69  & 92337.84  & 118833.32   \\
         \bottomrule
     \end{tabular}
     \end{adjustbox}
 \end{table}

\subsection{Discussion}

The analysis of stock price prediction using hybrid quantum-classical models highlights their advantages and limitations. The results consistently indicate that hybrid models outperform purely quantum models in predictive accuracy, with HybridQNN2 achieving the lowest RMSE across all configurations. The Gaussian error distribution and violin plot analysis further confirm that hybrid models exhibit lower variance and greater stability, reinforcing their effectiveness in mitigating prediction inconsistencies. 

However, all models struggle with rapid market transitions, underscoring the need for adaptive mechanisms to improve responsiveness to abrupt changes. A critical trade-off emerges between model complexity and computational cost, as increasing the qubit count enhances predictive accuracy but significantly prolongs training time, particularly for HybridQNN1, which incurs the highest computational overhead in the 5-qubit setting. This emphasizes the importance of efficient quantum circuit design to balance performance and resource consumption. 

Comparisons with classical deep learning models reveal that while hybrid models improve upon purely quantum approaches, they still fall short of the performance achieved by state-of-the-art architectures such as LSTM and BiLSTM in terms of RMSE. This suggests hybridization helps mitigate quantum limitations but does not yet confer a decisive advantage over classical models. Several directions for improvement emerge, including enhancing adaptability through reinforcement learning to better handle market fluctuations, optimizing hybrid architectures to reduce circuit depth and improve qubit utilization, exploring alternative quantum encoding schemes to enhance feature representation, and leveraging quantum error mitigation to address noise-related challenges and enhance model reliability. 

Overall, hybrid quantum-classical models represent a promising avenue for financial forecasting but require further optimization to compete effectively with classical deep learning approaches. Future advancements in quantum hardware, noise reduction techniques, and algorithmic refinements will be crucial in narrowing the performance gap and unlocking the full potential of quantum-enhanced financial models.

\section{Conclusion \label{sec5}}

This paper introduces a hybrid quantum-classical model for stock market forecasting, leveraging the strengths of both classical recurrent models and quantum circuits to capture complex financial time series patterns. The results demonstrate that integrating quantum models with classical architectures enhances predictive accuracy and robustness, particularly when compared to standalone quantum models. The analysis confirms that HybridQNN1 and HybridQNN2 consistently outperform CustomQNN, highlighting the effectiveness of parameterized rotation and entanglement layers in expanding the feature space. This increased dimensionality improves pattern recognition, especially under volatile market conditions, reinforcing the potential of QML in financial forecasting.

Interpretability played a key role in this study, as the SHAP analysis shows that domain-specific technical indicators, such as RSI and MACD, significantly influenced the model's predictive output. This emphasizes the importance of incorporating well-selected financial features to improve model performance. Additionally, HybridQNN2 exhibited a more balanced error distribution, reducing extreme prediction errors and enhancing stability, as reflected in the RMSE analysis.

Despite the advantages, several limitations remain. Quantum circuit execution is resource-intensive, and performance is constrained by the current state of Noisy Intermediate-Scale Quantum (NISQ) devices. Future work will explore larger-scale quantum architectures, including noise-resilient quantum processors, to enhance model scalability and precision. Furthermore, extending the model with larger datasets and additional financial indicators could improve its adaptability and generalization. Testing real-time applications on quantum hardware will also be crucial for assessing its practical feasibility in financial markets.

The key takeaway from this research is that hybrid quantum-classical models offer a promising approach to financial forecasting. While not yet surpassing classical deep learning methods, they provide a viable path toward quantum-enhanced predictive analytics. As quantum hardware advances and cloud-accessible QC matures, hybrid models can play a transformative role in financial decision-making, offering enhanced accuracy, interpretability, and deeper insights into complex market dynamics.

\section*{Acknowledgment}
This work was supported in parts by the NYUAD Center
for Quantum and Topological Systems (CQTS), funded by
Tamkeen under the NYUAD Research Institute grant CG008.

The authors also acknowledge the National Supercomputing Mission (NSM) for providing computing resources of 'PARAM Shivay' at the Indian Institute of Technology (BHU), Varanasi, which is implemented by C-DAC and supported by the Ministry of Electronics and Information Technology (MeitY) and Department of Science and Technology (DST), Government of India.


\bibliographystyle{IEEEtran}

\bibliography{refs}

\end{document}